\documentclass[usenatbib,useAMS,fleqn]{mnras}
\usepackage{color}
\usepackage[dvipsnames]{xcolor}
\usepackage{aas_macros}
\usepackage{graphicx}
\usepackage{upgreek}
\usepackage{amssymb}
\usepackage{amssymb,amsfonts,amsmath,times,pict2e}
\usepackage{graphicx}
\usepackage{enumerate}
\usepackage{subfigure}
\usepackage{bm}
\usepackage{amsmath,amsbsy}
\usepackage{underscore}
\usepackage{float} 
\usepackage{mathtools}
\usepackage{booktabs}
 \usepackage{romannum}

\newcommand{\fg}[1]{Fig. \ref{fig:#1}}
\newcommand{\Fg}[1]{Figure \ref{fig:#1}}%beginning of the sentence

\newcommand{\eq}[1]{Eq.~(\ref{eq:#1})}
\newcommand{\Eq}[1]{Equation~(\ref{eq:#1})}%beginning of the sentence

\newcommand{\tb}[1]{Table \ref{tab:#1}}
\newcommand{\Tb}[1]{Table \ref{tab:#1}}%beginning of the sentence
\newcommand{\tbs}[2]{Tables \ref{tab:#1} and \ref{tab:#2}}
\newcommand{\se}[1]{Sect.~\ref{sec:#1}}
\newcommand{\Se}[1]{Section~\ref{sec:#1}}%beginning of the sentence
\newcommand{\ses}[2]{Sects.\ \ref{sec:#1} and \ref{sec:#2}}

\newcommand{\ap}[1]{Appendix ~\ref{ap:#1}}
\newcommand{\eg}{e.g.}

\newcommand{\AU}{ \ \rm AU}
\newcommand{\rtran}{r_{\rm tran}}
\newcommand{\Mp}{M_{\rm p}}
\newcommand{\taus}{\tau_{\rm s}}
\newcommand{\taud}{\tau_{\rm dep}}
\newcommand{\alphat}{\alpha_{\rm t}}
\newcommand{\alphag}{\alpha_{\rm g}}
\newcommand{\Msyr}{  M_{\odot} \ \rm  yr^{-1}}
\newcommand{\dotMg}{  \dot M_{\rm g} }
\newcommand{\Meyr}{  M_{\oplus} \ \rm  yr^{-1}}
\newcommand{\dotMp}{  \dot M_{\rm peb} }
\newcommand{\Ms}{  M_{\star}}
\newcommand{\Msun}{  M_{\odot}}
\newcommand{\Me}{ \ M_{\oplus}}

\newcommand{\Rhill}{ R_{\rm H}}

\newcommand{\Ls}{L_{\star }}

\newcommand{\Mopt}{M_{\rm opt}}
\newcommand{\Miso}{M_{\rm iso}}
\newcommand{\Mgap}{M_{\rm gap}}

\title [Multiple, Distant Giant Planets Formation] {Promoted Mass Growth of Multiple, Distant  Giant Planets  through Pebble Accretion and Planet-Planet Collision} 
\author[John Wimarsson, Beibei Liu \& Masahiro Ogihara]{ John Wimarsson$^{1}$, Beibei Liu$^{1}$ \thanks{E-mail:bbliu@astro.lu.se  } and Masahiro Ogihara$^{2}$\\
$1$ Department of Astronomy and Theoretical Physics, Lund Observatory, Box 43, SE--221 00, Lund, Sweden.\\
$2$ Division of Science, National Astronomical Observatory of Japan, 2-21-1, Osawa, Mitaka, 181-8588 Tokyo, Japan}
\begin{document}

\date{Draft: \today}
\label{firstpage}
\pagerange{\pageref{firstpage}--\pageref{lastpage}}

\maketitle

\begin{abstract}

We propose a pebble-driven planet formation scenario to form giant planets with high multiplicity and large orbital distances in the early gas disk phase. We perform N-body simulations to investigate the growth and migration of low-mass protoplanets in the disk with inner viscously heated and outer stellar irradiated regions. The key feature of this model is that the giant planet cores grow rapidly by a combination of pebble accretion and planet-planet collisions. This consequently speeds up their gas accretion. Because of efficient growth, the planet transitions from rapid type I migration to slow type II migration early, reducing the inward migration substantially. Multiple giant planets can sequentially form in this way with increasing semimajor axes. Both mass growth and orbital retention are more pronounced when a large number of protoplanets are taken into account compared to the case of single planet growth. Eventually, a few numbers of giant planets form with orbital distances of a few to a few tens of AUs within $1.5{-}3$ Myr after the birth of the protoplanets. The resulting simulated planet populations could be linked to the substructures exhibited in disk observations as well as large orbital distance exoplanets observed in radial velocity and microlensing surveys.

\end{abstract}


\section{Introduction}

Giant planets are commonly observed in exoplanetary systems.  
Radial velocity surveys found that ${\sim}10{-}15\%$ of solar-type stars have gas giant planets with $M_{\rm p} {\gtrsim}50{-}100 \Me$, while most of them have orbital distances larger than $1$ AU \citep{Cumming2008,Mayor2011,Fernandes2019}. Furthermore, ${\sim}25{-}30\%$ of stars with a known giant planet host additional giant planet companions \citep{Wright2009,Wittenmyer2020}. Multiple gas giant planets are also likely to form around  metal-rich stars \citep{Buchhave2018}.
 Microlensing surveys reported that most abundant planets at orbital distances larger than a few AUs have Neptune to Saturn mass with a planet-to-star mass ratio peaked at the order of $ 10^{-4}$ \citep{Gould2010,Suzuki2016,Suzuki2018}.

On the other hand, planets can also be inferred from their fingerprints exhibited in natal protoplanetary disks. Recently, the Atacama Large Millimeter/submillimeter Array (ALMA) surveys have revealed the structures of young protoplanetary disks in great detail with unprecedentedly high sensitivity and angular resolution \citep{ALMA2015,Andrews2018}.
Axisymmetric rings and gaps are commonly seen among these disks \citep{Huang2018,Long2018}, in both early (HL Tau of ${\sim} 1$ Myr) and late evolved stages (TW Hydra of ${\sim}10$ Myr). Since such features are observed in both solid and gas components of disks \citep{Isella2016}, one promising interpretation is that these substructures are induced by the embedded planets through their interactions with disk gas \citep{Pinilla2012,Dipierro2015,Dong2015}. The corresponding planets deduced from the width/depth of the gaps from ALMA disk observations have Neptune to Jupiter mass \citep{Zhang2018,Bae2018,Lodato2019}. If this is the case, the ubiquitous nature of disk substructures inevitably implies that the formation of multiple giant planets beyond $10$ AU is very efficient, even in the early gas-rich disk phase. 

Based on the core accretion scenario, the giant planet should firstly assemble solids to form a sufficiently massive core in the gas-rich protoplanetary disk \citep{Pollack1996}. Then the planet can subsequently accrete surrounding gas before the disk gas is entirely depleted (${\sim}3{-}10$ Myr, \cite{Haisch2001}). 
Forming gas giant planets at radial distances of tens of AU from the central star is challenging in classical planetesimal-driven planet formation scenarios  \citep{Ida2004a,Goldreich2004}.  Alternatively, \cite{Lambrechts2014} firstly proposed that the growth by pebble accretion  \citep{Ormel2010,Lambrechts2012} in the outer disk can be sufficiently fast to form giant planets.

\cite{Bitsch2015} found that Jupiter-like, cold gas giant planet can form by pebble accretion when the initial protoplanet is born at a disk location between $20$ AU to $40$ AU, since the planet experiences significant inward orbital migration. 
% \footnote{ \cite{Pirani2019} reported that such inward migration and planet growth naturally produce asymmetry feature of Jupiter's Trojans.}. 
They only considered the growth of one single protoplanet.  \cite{Liu2015} investigated the dynamical evolution of multiple protoplanets where they migrate and get trapped at a transition radius that separates two disk heating mechanisms. Such a concentration of planets leads to planet-planet collisions, significantly promoting the formation of massive cores. Further, other similar studies purely focus on the growth of planetary cores by collisions among low-mass protoplanets \citep{Cossou2014,Coleman2014,Ogihara2015,Izidoro2017,Ogihara2018}.   

Here, we are interested in exploring the planet growth by a combination of pebble accretion and planet-planet collisions. In such a case, the core  growth rate is largely enhanced by these two processes. Motivated by the aforementioned observations, we provide a scenario where multiple giant planets can form in the early disk phase at large orbital distances. The key properties of the giant planets we aim to  investigate is their formation time, location and multiplicity. 

The paper is structured as follows. We describe the model in \se{method}. The illustrated simulations are presented in \se{result} and a parameter study is conducted in \se{parameter}. Finally, we discuss our results in \Se{discussion}  and summarise the conclusions in \se{conclusion}.

\section{Method}
\label{sec:method}
In this paper we adopt the pebble-driven planet formation model from \cite{Liu2019b}.  Detailed descriptions of the model can be found in their Section 2. In this section we recapitulate important equations and highlight the key features of this model.

\subsection{Disk model}
\label{sec:disk}

The adopted protoplanetary disk has two components, an inner viscously heated region and an outer stellar irradiated region. The gas surface density and disk aspect ratio are given by 
  \begin{equation}
 \frac{\Sigma_{\rm g}}{ \rm g \ cm^{-2}} { =}
  \begin{cases}
 {\displaystyle   235
    \left( \frac{\dot M_{\rm g}}{10^{-7} \Msyr} \right)^{1/2}  \left(\frac{M_{\star}}{ M_{\odot}} \right)^{1/8}
    } \\
      {\displaystyle  \left(\frac{\alpha_{\rm g}}{10^{-2}} \right)^{-3/4} \left(\frac{\kappa_0}{0.1} \right)^{-1/4} \left(\frac{r}{ \AU} \right)^{-3/8}  } 
     \hfill  [\mbox{vis}],  \vspace{0.1cm}\\
       \hspace*{1 mm} \\ 
 {\displaystyle  2500 \left( \frac{\dot M_{\rm g}}{10^{-7} \rm \Msyr} \right)
\left(\frac{M_{\star}}{ M_{\odot}} \right)^{9/14} } \\
{\displaystyle  \left(\frac{L_{\star}}{ L_{\odot}} \right)^{-2/7}   \left(\frac{\alpha_{\rm g}}{10^{-2}} \right)^{-1}  \left(\frac{r}{ \AU} \right)^{-15/14} }
  \hfill  [\mbox{irr}], 
\end{cases}
\label{eq:sigma}
\end{equation}
and
  \begin{equation}
h_{\rm g} = \begin{cases}
 {\displaystyle   0.08
\left( \frac{\dot M_{\rm g}}{10^{-7} \Msyr} \right)^{1/4}
\left(\frac{M_{\star}}{ M_{\odot}} \right)^{-5/16} } \\
  {\displaystyle    \left(\frac{\alpha_{\rm g}}{10^{-2}} \right)^{-1/8}  \left(\frac{\kappa_0}{0.1} \right)^{1/8}  \left(\frac{r}{ \AU} \right)^{-1/16} } 
  \hfill  [\mbox{vis}],   \vspace{ 0.1cm}\\
      \hspace*{1 mm} \\ 
 {\displaystyle   0.0245
    \left(\frac{M_{\star}}{ M_{\odot}} \right)^{-4/7} 
    \left(\frac{L_{\star}}{L_{\odot}} \right)^{1/7}    \left(\frac{r}{ \AU} \right)^{2/7} }     \hfill  [\mbox{irr}].
\end{cases}
\label{eq:aspect}
\end{equation}
The separation between these two disk regions with different heating mechanisms is defined as the transition radius, 
\begin{equation}
      \begin{split}
  r_{\rm tran} = & 30
    \left( \frac{\dot M_{\rm g}}{10^{-7} \ \Msyr} \right)^{0.72}
    \left(\frac{M_{\star}}{ M_{\odot}} \right)^{0.74}  \left(\frac{L_{\star}}{ L_{\odot}} \right)^{-0.41} \\
    &  \left(\frac{\alpha_{\rm g}}{10^{-2}} \right)^{-0.36}  \left(\frac{\kappa_0}{0.1} \right)^{0.36}   \AU.
     \end{split}
     \label{eq:rtrans}
\end{equation}
In above equations\footnote{  See Eq. (8)-(14) of \cite{Liu2019b} for comparison. }, $\dotMg$, $\Ms$, $\Ls$ and $r$ are the disk gas accretion rate, the stellar mass, the stellar luminosity and the disk radial distance to the central star, respectively. 
We assume a steady state disk such that  ${\dot M}_{\rm g} {= } 3 \pi \Sigma_{\rm g} \alphag h_{\rm g}^2 r^2 \Omega_{\rm K}$, 
where $\Omega_{\rm K}$ is the Keplerian angular velocity and $\alphag$ corresponds to the global angular momentum transfer efficiency. The value of  $\alphag$  is adopted to be $10^{-2}$, motivated from disk observations \citep{Hartmann1998}. We also define  the dimensionless  parameter of the local turbulent viscosity or the coefficient of local gas diffusivity as $\alpha_{\rm t}$. %The detailed reasons for two $\alpha$ scheme and the motivation of specific  values of $\alpha_{\rm g}$ and $\alpha_{\rm t}$ can be referred to Section 2 of \cite{Liu2019b}.

The physical motivation for this two-$\alpha$ scheme is based on a layered accretion assumption where the disk is envisioned by a thin, quiescent midplane region and an active, turbulent zone upwards    \citep{Gammie1996,Fleming2003}. The value of $\alphat$ characterizes the weak turbulent diffusion in the midplane of the dead zone while $\alphag$ represents the vertically averaged, angular momentum transfer efficiency, which is dominated by stronger turbulence in the active zone. As shown by \cite{Bitsch2015b}, a disk with this structure is essentially vertically isothermal and its temperature is very close to that of a disk with a single value of $\alphag$ and the same accretion rate onto the star. This justifies the radial dependence of the temperature described in \cite{Liu2019b}.

 We note that our adopted disk is a classical viscous accretion driven model, which does not account for Magnetohydrodynamics (MHD) disk winds \citep{Bai2013,Gressel2015,Bethune2017}. In those studies when other non-ideal MHD effects such as ambipolar diffusion are included,  magnetorotational instability (MRI) is quenched even in the active layer and the accretion occurs at the surface of the disk due to angular momentum removal by the magnetocentrifugal wind. Despite that, the detailed outcomes depend on the geometry and strength of the magnetic field. \cite{Mori2019} performed non-ideal MHD simulations and showed that the midplane temperature derived from their simulations is lower than the temperature in our model due to the lack of viscous heating. Our model cannot be applicable in those circumstances.  The inner viscously heated disk is warm in our model during the early gas-rich disk phase, and therefore, the transition radius is far from the central star. 

In this work and \cite{Liu2019b}, the disk opacity is given by $\kappa {=} \kappa_0 (T/1 \rm \ K) \ cm^{2}g^{-1}$, where $T$ is the  gas temperature. The opacity coefficient $\kappa_0$ is adopted to be $10^{-2}$ in \cite{Liu2019b}.  \cite{Bell1994} provided a opacity expression of $\kappa {=} 2{\times} 10^{-4} (T/1 \rm \ K)^2 \ cm^{2}g^{-1}$ beyond the water ice line. The resultant dust opacity of \cite{Liu2019b} at $r{=}30 $ AU ($T{=} 35 \rm \ K$) is $0.35 \rm \ cm^{2}g^{-1}$, comparable to \cite{Bell1994}'s value of $0.25 \rm \ cm^{2}/g$ at the same temperature, although these two opacity expressions have different temperature scaling.  However, we note that \cite{Bell1994}'s opacity calculation is based on the assumption that  grains are compact spheres, following a ISM-like size distribution. 
The porous dust aggregates would have much higher opacity than the dust grains with compact structures due to a higher area-to-mass ratio.  A realistic value and temperature scaling for the disk opacity remains unknown, especially when considering the detailed dust coagulation with different chemical compositions and porosity evolution. 
Here,  we  choose  $\kappa_0 {=} 0.1$, resulting in  a high opacity  of $3.5 \rm \ cm^{2}g^{-1}$ at $30$ AU. The adopted opacity value and temperature scaling is more in line with \cite{Semenov2003}'s iron-deficit, porous composite spherical grain model (see the left panel of their Fig.1).  In this circumstance, the disk is more opaque and has a larger $r_{\rm tran}$ than in \cite{Liu2019b} or when adopting \cite{Bell1994}'s opacity law. For comparison, $r_{\rm tran} {=}30$ AU in this work, while $r_{\rm tran} {\simeq}13$ AU if the disk opacity efficiency is replaced by \cite{Liu2019b}'s value or when using \cite{Bell1994}'s expression beyond the water-ice line.  The choice of a high opacity here facilitates the aim of the paper, to explore the formation of distant giant planets.
 
 We assume that in the early gas-rich disk phase,  the disk accretion rate remains a constant. After a time $t_0$, the disk gas dissipates exponentially with a timescale of $\taud$. The time-evolution of the disk accretion rate can therefore be written as  
 \begin{equation}
{\dot M}_{\rm g} =\begin{cases}
 {\dot M}_{\rm g0}  & \mbox{ when $ t \leq t_0$ }, \\
 {\dot M}_{\rm g0}   \exp \left[- (t-t_0)/\taud \right] & \mbox{ when $ t > t_0$ }. \\
 \end{cases}
\label{eq:taudep}
\end{equation}
In order to explore the early formation of giant planets, we assume the disk maintains a relatively high accretion rate at an early time. The fiducial parameters are set as  $M_{\rm g0} {= }10^{-7} \Msyr$, $t_0{=}1$ Myr and  $\taud{=}0.5$ Myr.  The total disk mass in this case is $0.15 \Msun$.  Such early, high accretion disks are essential  in our model. As we will show in \se{multiple}, the planets in these circumstances can undergo fast convergent migrations and results in planet-planet collisions. 
  
  \subsection{Planet growth and migration} 
  \label{sec:growth}
The growth of protoplanets include pebble accretion onto cores and gas accretion onto envelopes.   
For pebble accretion, the mass growth rate  is given by 
  \begin{equation}
 \begin{split}
 \frac{\dot M_{\rm p, peb}}{ \rm \Meyr}   
 & { =}     \varepsilon_{\rm PA} \dotMp \\
 & = \begin{cases}
 {\displaystyle   5{\times}10^{-7}
    \left( \frac{ \dotMp}{10^{-4} \Me \rm \ yr^{-1}} \right)  \left(\frac{M_{\rm p}}{ 0.05 \Me} \right)^{2/3} }\\
    {\displaystyle \left(\frac{\taus}{ 10^{-2}} \right)^{-1/3} \left(\frac{\eta}{ 5.5{\times}10^{-3}} \right)^{-1} }    \hspace*{1.7cm}   [\mbox{2D}]  \\
    \vspace{0.1cm} \\
    {\displaystyle  2{\times}10^{-7} \left( \frac{\dotMp}{10^{-4} \rm \Me \rm \ yr^{-1}} \right)
\left(\frac{M_{\rm p}}{ 0.05 \Me} \right)  } \\
{\displaystyle \left(\frac{h_{\rm peb}}{ 6.5{\times}10^{-3}} \right)^{-1}  \left(\frac{\eta}{5.5 {\times} 10^{-3} } \right)^{-1} }
  \hspace*{1.2cm}   [\mbox{3D}], 
\end{cases}
\label{eq:Mpeb}
 \end{split}
\end{equation}
where $\dotMp$ is the disk pebble flux, $\varepsilon_{\rm PA}$ is the pebble accretion efficiency, $\Mp$ is the planet mass, $\taus$ is the Stokes number of pebbles, $h_{\rm peb}$ is the aspect ratio of the pebble disk,  $\eta$ measures the relative difference between the gas azimuthal velocity and the Keplerian velocity $v_{\rm K}$. For illustration,  the upper panel of  the above equation only depicts the accretion in the shear dominated regime  where the relative velocity between planet and pebble is dominated by the Keplerian shear velocity \citep{Lambrechts2014,Liu2018}.  Whether the accretion is in the $2$D/$3$D regime is determined by the ratio of the accretion radius of the planet and the pebble scale height \citep{Morbidelli2015, Ormel2018}. The growth is in $2$D (3D) when this ratio is larger (smaller) than unity.
We note that the pebble  accretion prescription used in this work  considers both $3$D accretion and headwind/shear-dominated $2$D accretion. The effect of eccentricity and inclination of the planet's orbit is also taken into account (see  details in \citealt{Ormel2018}).

There are two key parameters that regulate the core accretion rate: the pebble mass flux $\dotMp$  (the total mass of the pebble reservoir) and the Stokes number of pebbles $\taus$ (the aerodynamical size of pebbles). Here, we set the pebble mass flux $\dot M_{\rm peb} {=}1.5{\times}10^{-4} \Meyr$ at the beginning and it follows the same time-evolution as the disk accretion rate  (\eq{taudep}). In turn, this means that $0.3\%$ of the total disk mass is in pebbles ($ M_{\rm peb}/ M_{\rm d} {=}\dot M_{\rm peb}/\dot M_{\rm g}{=}0.3\%$). We also assume a constant Stokes number of the pebbles, $\taus {=}10^{-2}$. The influence of these two parameters on the evolution of the planet growth will be discussed in \se{parameter}. It is worth noting that we need to assume overall large size disks, and particularly very extended disks (\eg , ${>}500$ AU) for the high Stokes number case in \se{pebble} in order to maintain such a constant flux ratio for a few Myr. The disk size is much larger compared to the planet formation region under investigation. Such disk sizes may be difficult to find in a few Myr old systems due to the rapid radial drift of pebbles. Nevertheless, disk sizes are larger when they are younger. For example, 
$\rm IM \  Lup$ \citep{Cleeves2016} is a very young system with an extended gas/dust disk of a few hundred AU.

We assume that the gas accretion occurs when the core reaches pebble isolation mass (see \eq{miso} below). The corresponding accretion rate onto the planet envelope is given by 
  \begin{equation}
 \dot M_{\rm p, g} = \min \left[ \left(\frac{d M_{\rm p, g}}{dt}\right)_{\rm KH} ,\left(\frac{d M_{\rm p, g}}{dt}\right)_{\rm Hill}, \dot M_{\rm g}  \right].
 \label{eq:gas}
 \end{equation} 
The first term on the right hand side of \eq{gas} represents the Kelvin-Helmholtz contraction based on \cite{Ikoma2000}, 
\begin{equation}
 \left(\frac{d M_{\rm p, g}}{dt}\right)_{\rm KH} =10^{-5} \left( \frac{\Mp }{10 \Me} \right)^4 \left( \frac{\kappa_{\rm env}}{1 \rm \ cm^2g^{-1}}  \right)^{-1} \rm \  M_{\oplus}\,yr^{-1},
\label{eq:gasKH}
 \end{equation}
 where $\kappa_{\rm env}$ is the opacity in planetary envelope.  
The second term in \eq{gas} sets how much gas can be accreted in the planet's Hill sphere \citep{Liu2019b}, 
  \begin{equation}
  \begin{split}
\left(\frac{d M_{\rm p,  g}}{dt}\right)_{\rm Hill}&  =  0.02 \left(\frac{M_{\rm p}}{10  \Me} \right)^{2/3} \left(\frac{\dotMg}{10^{-7} \Msyr} \right)   \\
&   \left(\frac{h_{\rm g}}{0.065} \right)^{-2} \left[ 1 + \left(\frac{M_{\rm p}}{\Mgap }\right)^2 \right]^{-1} \Meyr,
\label{eq:gashill}
  \end{split}
\end{equation}
where $\Mgap$ is the gap opening mass (see \eq{mgap} below).  
We note that $\kappa_{\rm env}$  is a key free parameter that determines the amount of gas accreted onto the planet. 
We test different $\kappa_{\rm env}$ and find that a very low $\kappa_{\rm env}{<} 0.1 \rm \ cm^{2}g^{-1} $ results in too many, very massive giant planets, which conflicts with observations. On the other hand, the disk opacity calculated from theory is roughly order of unity beyond the water-ice line, depending on the grain size and abundance. Since  grain sedimentation might occur in the planet envelope,  the envelope opacity should be at least no higher than the disk opacity.  Considering the above two aspects, we choose a moderate value of $\kappa_{\rm env}{=}0.5 \rm \ cm^{2}g^{-1}$.

Planets gravitationally interact with their natal protoplanetary disks, leading to orbital migration (see \citet{Kley2012} for a review).  Low-mass planets undergo type I migration whereas massive giant planets are in the type II regime. 
The transition planet mass from type I to type II is defined as the gap opening mass, which is given by \citep{Kanagawa2015}, 
\begin{equation}
M_{\rm gap} =  18 \left(  \frac{\alphat}{10^{-4}} \right)^{1/2} \left( \frac{h_{\rm g}}{0.065} \right)^{5/2} \left( \frac{M_{\star}}{\Msun}  \right) \Me,
\label{eq:mgap}
\end{equation}
where $\alphat$ is the coefficient of local gas diffusivity, representing the disk turbulent level. We adopt $\alphat{=}10^{-4}$ in this work, motivated by the $\rm CO$ line broadening measurements \citep{Flaherty2015}.     

The pebble isolation mass is also an important quantity. It refers to a planet that is massive enough to open a shallow gap and induce a local pressure maximum in its vicinity. In such a circumstance, the inward drifting pebbles stop at the local pressure maximum and cannot be accreted by the planet.
Therefore, the core mass growth is quenched when the planet reaches the pebble isolation mass \citep{Lambrechts2014b}. However, in order to produce a deep enough gap to slow down the planet migration the planet needs to reach gap opening mass. In principle, the pebble isolation mass should be lower than the gap opening mass.
   
\cite{Johansen2019} simulated a planet  embedded in a $1$ D disk with torque formulas adopted from \cite{D'Angelo2010} and they found that the gap opening mass is $2.3$ times larger than the pebble isolation mass. They define these two quantities as when the gap depths are reduced by $50\%$ and $15\%$, respectively (see their Fig. 3). 
Nevertheless, the pebble isolation mass could vary in $2/3$ D compared to $1$ D \citep{Bitsch2018,Ataiee2018}.

In this work we conveniently adopt the pebble isolation mass to be $2.3$ times lower than the gap opening mass,    
\begin{equation}
M_{\rm iso} =  8 \left(  \frac{\alphat}{10^{-4}} \right)^{1/2} \left( \frac{h_{\rm g}}{0.065} \right)^{5/2} \left( \frac{M_{\star}}{\Msun}  \right) \Me.
\label{eq:miso}
\end{equation}

The total migration torque is expressed as 
\begin{equation}
  \Gamma {= }\left[ f_{\rm I} f_{\rm s}  +  f_{\rm II}  (1-f_{\rm s}) \right] \Gamma_0
\label{eq:torque}
\end{equation}
where   $\Gamma_0 {=}M_{\rm p}^2 \Sigma_{\rm g} r^{4} \Omega_{\rm K}^{2}/M_{\star}^2h_{\rm g}^{2}$ is the normilized type I torque, $f_{\rm I}$ is the type I migration coefficient adopted from \cite{Paardekooper2011},  and $f_{\rm II}{=} -1/ \left(M_{\rm p}/ \Mgap \right)^2 $ is the reduced type II migration coefficient based on \cite{Kanagawa2018}. We use a smoothing function  $f_{\rm s} {=} \exp[-(M_{\rm p}-\Mgap)/\Delta M]$ to combine these two regimes, where $\Delta M {=}0.2\Mgap$. We note that the formula of eccentricity and inclination damping is adopted from \cite{Cresswell2008}. The detailed prescription of  the type I  torque ($f_{\rm I} \Gamma_0$) is given in \ap{torque}. 
We note that we neglect  the effects of dynamical torques \citep{Paardekooper2014} in this work, as well as the additional gap-deepening due to planet gas accretion \citep{Crida2017}, which might play certain roles in reducing the migration rate.

Due to the unsaturated nature of the corotation torque \citep{Paardekooper2011}, we define another important planet mass as the optimal mass, 
\begin{equation}
M_{\rm opt} = 2\left(\frac{\alphat} {10^{-4}} \right)^{2/3}\left(\frac {h_{\rm g}}{0.065} \right)^{7/3}\left(\frac{M_{\star}}{1 \ M_{\odot}} \right) \ M_{\oplus}.
\label{eq:mopt}
\end{equation}
When $M_{\rm p} {\simeq} M_{\rm opt}$, the planet can undergo outward type I migration to the transition radius in the viscously heated disk region, while the planet directly migrates inward when $M_{\rm p} {\gg} M_{\rm opt}$ \citep{Kretke2012,Liu2015}. 
We calculate the radius of the planet from a mass-radius relation based on the fitting of solar system planets \citep{Lissauer2011}, $R_{\rm p}/R_{\oplus} = (M_{\rm p} / \Me)^{1/2.06}$.

We use a two $\alpha$ parameter approach in this study. The global $\alphag$ sets  the disk angular momentum transportation and large scale disk structure  (\eg , $\Sigma_{\rm g}$, $h_{\rm g}$) while $\alphat$ is more relevant to local planet formation processes (planet gap opening, gas diffusion across the gap) occurring in the disk midplane.  The caveat is that the gap opening mass, the isolation mass, \cite{Paardekooper2011}’s torques and the optimal mass (Eqs. 9, 10, 12 and Appendix A) are obtained from hydrodynamical simulations based on a single viscosity approach. Whether and how the gap-opening and corotation saturation would change in the layered accretion disks is not yet well understood. Although we neglect the influence of viscous accretion in the upper layer on the relevant processes occurred in the disk midplane for the sake of simplicity, we admit that our model needs to be updated in future when more results are reported from dedicated hydrodynamical simulations on this topic.

\section{Results}
\label{sec:result}

In this work, we start from protoplanets that are assumed to form by streaming instability \citep{Youdin2005}, where dust particles are clustered and directly collapses into planetesimals. The mass distribution of the forming planetesimals can be fitted by a power-law plus an exponential decay \citep{Johansen2015,Schafer2017}. 
Based on the extrapolation of literature streaming instability simulation studies, \cite{Liu2020} derived the characteristic planetesimal mass (their Eq. 13) and the mass of the largest bodies from the forming planetesimal population (named protplanets, their Eq. 14). The mass of the protoplanet is roughly $0.1 \Me$ at $30$ AU  for our adopted disk model. Since the protoplanets will dominate the following mass growth and dynamical evolution of the whole population, we only focus on them hereafter. The masses of the protoplanets are all assumed to be $0.05 \Me$ for simplicity. The starting time $t{=}0$ yr in our study is the birth time of these protoplanets, which approximately is the onset time of streaming instability in protoplanet disks.

We conduct numerical simulations to investigate the growth and migration of protoplanets around a solar-mass star. We use the {\it Mercury} N-body code \citep{Chambers1999} and adopt the Bulirsch-Stoer integrator.  In addition, the code includes the effects of pebble accretion, gas accretion, type I and type II migration and gas damping.  Perfect merger is treated here such that when the separation of two planets is smaller than the sum of their physical radii, they collide into one with conservation of mass and angular momentum.  We simulate over $5$ Myr until the disk is fully depleted. The long-term secular evolution of planets after  disk dispersal is not taken into account in this work. The subsequent evolution of planetary systems in gas-free environment will be investigated in a future study. 
The results for the growth of one single protoplanet and multiple protoplanets are presented in \se{single}  and \se{multiple}, respectively.

\subsection{Growth of single protoplanet}
\label{sec:single}

\begin{figure}
 \includegraphics[scale=0.46, angle=0]{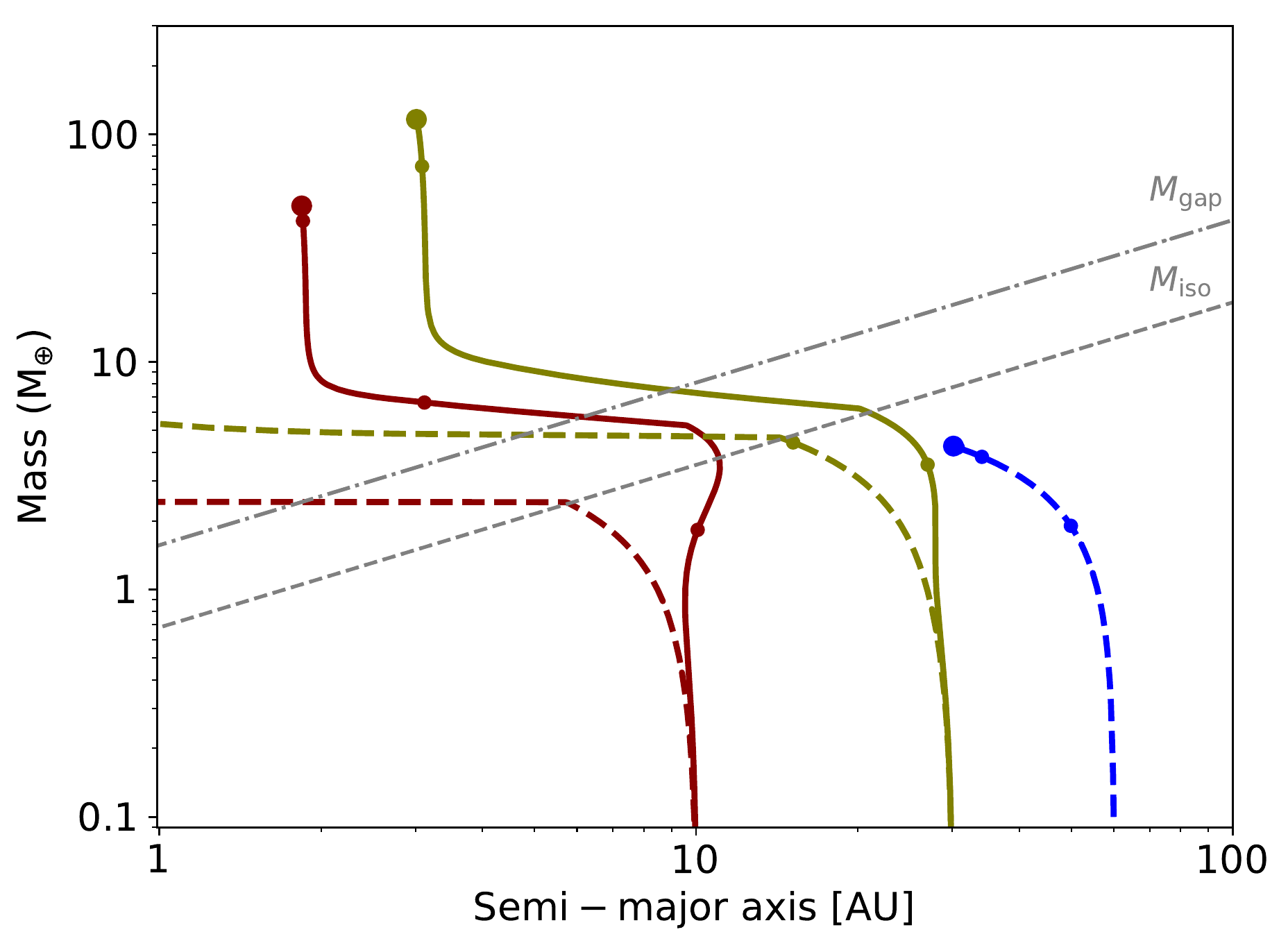}
       \caption{ Growth track (mass vs semimajor axis) of single protoplanet in a purely stellar irradiated disk (dashed) and in a two-component disk with an inner viscously heated region and an outer stellar irradiation region (solid). The dashed and dotted-dashed lines refer to the pebble isolation mass and the gap opening mass in a purely stellar irradiated disk.   The initial locations of the protoplanets are at $10$ AU (red), $30$ AU ($=\rtran$, yellow) and $60$ AU (blue). The initial mass of the protoplanet is assumed to be $0.05 \Me$.   The small circles mark at every $1$ Myr  and large circles indicate at the disk evolution time of $5$ Myr. For specific growth traces, we see  these circles and dots overlap with each other since planets seldom grow and migrate  after $3{-}4$ Myr. The corresponding migration map can be referred to \fg{map} for comparison.
    }
\label{fig:mr}
\end{figure}

\Fg{mr} illustrates the growth tracks of individual protoplanets starting at different disk locations, either in a purely stellar irradiated disk (dashed) or in a two-component disk with an inner viscously heated region and an outer stellar irradiated region (solid). The grey dash-dotted and dashed lines refer to $\Mgap$ and $\Miso$ in the stellar irradiated disk.  

Planets never migrate outward in the stellar irradiated disk. When the protoplanets  originate  at  $10$ AU (red) and $30$ AU (yellow), they undergo rapid type I inward migration and attain their pebble isolation masses at $2\Me$ and $4 \Me$  before reaching the most inner region of the disk. When the protoplanet forms at $60$ AU (blue), it cannot reach the pebble isolation within the disk lifetime, and finally grows into a $3\Me$ planet. As shown in \fg{mr}, protoplanets maximally grow to core-dominated super-Earths without any significant gaseous envelopes in a disk that is purely heated by stellar irradiation.

Nevertheless, planet growth  has two advantages in the disk with an inner viscously heated region and an outer stellar irradiated region compared to that in the purely stellar irradiated disk. First, before migrating inward substantially, planets of a few Earth masses ($\Mp {\sim} \Mopt$, see \fg{map}) can already migrate outward to the transition radius $\rtran$, which is located at $30$ AU.  Therefore, they can remain at distant disk locations for a longer time in a two-component disk as they will not directly migrate inward, which is the case for a purely stellar irradiated disk.  Second, in the inner viscously heated disk region, $h_{\rm g}$ is almost  independent of $r$ (\eq{aspect}), and therefore $\Miso$ remains roughly a constant of $ 8 \Me$. Thus, in this case, the planet retained at $\rtran$ can reach a higher $\Miso$ compared to a planet in the purely stellar irradiated disk. Furthermore, planets of higher masses accrete the surrounding gas more rapidly.  Combining the two effects mentioned above, protoplanets can trigger runaway gas accretion to form giant planets when they reach isolation mass at $10$ AU and $30$ AU in a two-component disk. 

Since $\Mopt {<}\Miso$, the planets that have reached pebble isolation mass cannot stay at $\rtran$ but migrate inward (\fg{mr}). On the other hand, these migrating planets further accrete gas and gradually transition from fast type I to  slow type II migration when their masses become higher than $\Mgap$.  Therefore, the protoplanets that originate at $10$ AU and $30$ AU both migrate inward substantially, and finally grow into $50{-}100\Me$ gas giant planets at orbital distances of $2{-}3$ AU. 

To summarise, planets can migrate outward in the viscously heated disk region while they migrate inward in the stellar irradiated disk region. Thus, planets migrate toward and temporarily stay at the transition radius. This process both promotes the planet growth and reduces the inward migration, which  is of great importance for forming massive giant planets at large orbital distances.

\subsection{Growth of multiple protoplanets}
\label{sec:multiple}

\subsubsection{Illustrated simulation}

Here we study the growth and migration of multiple protoplanets.  All the disk (two-component disk model) and protoplanet properties (initial masses) are adopted to be the same as \se{single}. Initially, fifteen protoplanets are distributed around the transition radius from $16$ AU to $43$ AU with a separation of $15 \  \Rhill$, where $\Rhill {=}(2\Mp/3\Ms)^{1/3} a_{\rm p}$ is the mutual Hill radius and $a_{\rm p}$ is the planet semimajor axis. The eccentricities and inclinations of the protoplanet orbits are assumed to follow Rayleigh distributions where $e_0{=}2i_0{=}10^{-2}$ are the corresponding scale parameters.

\fg{example} illustrates the evolution of the masses and semimajor axes  of these protoplanets. The dashed yellow line corresponds to the case of a single protoplanet with its origin at $30$ AU (the same as solid yellow line in \fg{mr}) for comparison, and the thick cyan line represents $\rtran$.  At the beginning, the protoplanets slowly grow their core masses by pebble accretion, and they reach Earth-mass within $1$ Myr. Such planets have  already undergone type I migration towards $\rtran$. Since the migrations are convergent, the orbital spacing between planets gradually decreases. Dynamical interactions are further enhanced due to growing masses. These planets trapped at $r_{\rm tran}$ frequently overlap orbits with their neighbouring siblings. After a few $10^{5}$ yr, planets that undergo repeated close-encounters  eventually collide with each other.  When two planets merge into one, the mass of the new planet is the sum of the previous two bodies. Due to the fact that the pebble accretion efficiency increases with the planet mass (\eq{Mpeb}), the growth of this new planet is boosted by pebble accretion.

As can be seen in \fg{example}, initially, due to the influence of dynamical interactions among these protoplanets, the growth by pebble accretion (grey) is slower compared to the case of a single protoplanet (yellow dashed) when it is in a circular and coplanar orbit. However, the growth significantly speeds up when successive planet-planet collisions occur.  We see in \fg{example} that after two collisions one massive planet with $\Mp{>} 10 \Me$ form at $t{=}1.6$ Myr. 

This planet of $M_{\rm p}{>}\Miso$ can initiate a rapidly gas accretion. Meanwhile, it  gradually transitions from rapid type I migration into slow type II migration. The growth and migration of the massive planet by presence of multiple protoplanets has two key differences compared to that of the single protoplanet.  First, the fast inward migration of the massive planet could be slowed down by the inner lower-mass protoplanets.  Second and most importantly, the planet has a higher mass core due to collisions and thus accretes gas faster. Thus, this massive planet migrates more slowly in the type II regime due to its higher mass.  We find that a gas giant planet of $0.5 M_{\rm Jup}$ finally forms at $4.6$ AU (red). The main point here is that both the mass growth and orbital retention are more significant in this case compared to the case of a single protoplanet.  

In addition to the formation of one gas giant planet, we also find that such a convergent migration scenario is prone to form multiple, wide orbit giant planets. We find in \fg{example} that when the first gas giant is growing and migrating toward to  $4.6$ AU, the second massive planet (blue) forms outside the orbit of the first gas giant. This is because the strong gravitational perturbation from the giant planet promotes subsequent planet-planet scatterings/collisions. Furthermore, the rapid inward migration is largely reduced  by the inner slowly migrating gas giant. As a result, the second giant planet finally forms further out at $8.0$ AU. Sequentially, other massive planets grow in a similar way and end up at orbits exterior to inner neighbouring giant planets.  In total five gas giant planets form in \fg{example}, whose orbits are $4.6$ AU, $8.0$ AU, $13.4$ AU, $18.1$ AU  and $24.6$ AU, respectively, where the outer most three planets are trapped into $3$:$2$ mean motion resonances. 

The eccentricity  and inclination evolution of the planets are also illustrated in \fg{example}.  Initially, type I  torque damps the random velocities of the protoplanets. As these protoplanets grow and migrate towards a more compact configuration, they collide with each other to form massive planets. The mass difference among these protoplanets increases with time. Through mutual interaction and scattering processes, the eccentricities/inclinations of massive planets remain low, whereas  the random velocities of less massive planets are excited. 

\subsubsection{Observational connection}

We further discuss how our simulations could be linked to disk observations. ALMA surveys have commonly found the existence of substructures in young protoplanetary disks, which can be inferred due to the presence of multiple planets.   \cite{Eriksson2020} conducted $1$ D dust drift and coagulation model including already formed multiple giant planets. They found that when these planets are below the pebble isolation mass, the drifting pebbles can partially bypass the orbits of the planets, leading to multiple rings and gaps. In such a case, there are clear depletions of particles between neighbouring planets and slight concentrations of particles outside the orbits of the planets. On the other hand, when the planets are more massive than the pebble isolation mass,   all the drifting pebbles are blocked beyond the outmost giant planet. In this case, a large inner cavity of a few tens of AUs would be observed instead. 

 In our disk model, $1{-}3$ Myr old disks (light red region in \fg{example}) have accretion rates of $10^{-7}$ to $2{\times} 10^{-9} \Msyr$, matching the  typical observed $\dotMg$ for T Tauri stars. The progenitors of these gas giant planets have masses from $2\Me$ to $200 \Me$ at $1{-}3$ Myr and  semimajor axes range from $5$ to $40$ AU (\fg{example}). We speculate that planets could form slightly further out in disks with a higher $\dot {M}_{\rm g0}$ and therefore a larger $r_{\rm tran}$ (\eq{rtrans}). Nevertheless, the model can hardly generate giant planets at very large orbital distances (see \fg{parameter}). In fact, some of rings and gaps observed from DSHARP surveys are much beyond $50$ AU, which cannot explain by our model.  We discuss this further in \se{escape}. 
   
For our simulations, planets are below the pebble isolation mass at $t {\lesssim}1{-}1.5$ Myr. Rings and gaps can be formed in such circumstances. When the planet grows beyond the pebble isolation mass  ($t{\sim}2{-}3$ Myr),  the inner cavity is produced inside of the giant planets. It is worth mentioning that the formation time of giant planets also varies with different parameters (see \se{parameter}), resulting in observed substructures in disks  of various ages.  
 
 Another feature of our model is the formation of multiple giant planets. For instance, there are four giant planets, Jupiter, Saturn, Uranus and Neptune in our Solar System. The orbital range of these four planets is consistent with the simulated planets shown in \fg{example}.
We note that a super-Earth also forms inside of the orbit of the most inner giant planet. On the other hand, we did not assume any protoplanet interior of  $15$ AU. If such protoplanets had been considered, multiple super-Earths might have formed at the most inner disk region. After all, our goal here is not to reproduce the architecture of the Solar System. The key point we want to emphasize is that this convergent migration plus planet-planet collisions is a potential channel to grow multiple giant planets.
Observationally, the occurrence rate of giant planets is ${\sim} 10\%$ \citep{Cumming2008,Mayor2011,Fernandes2019}, whereas ${\sim}30\%$ of planetary systems with known giant planet contain additional companion(s) \citep{Wright2009,Wittenmyer2020}. It plausibly indicates that multiple giant planets are likely to form together once disks are massive enough to produce one giant planet, which is  in agreement with the picture we demonstrate here.

\begin{figure*}
 \includegraphics[scale=0.6, angle=0]{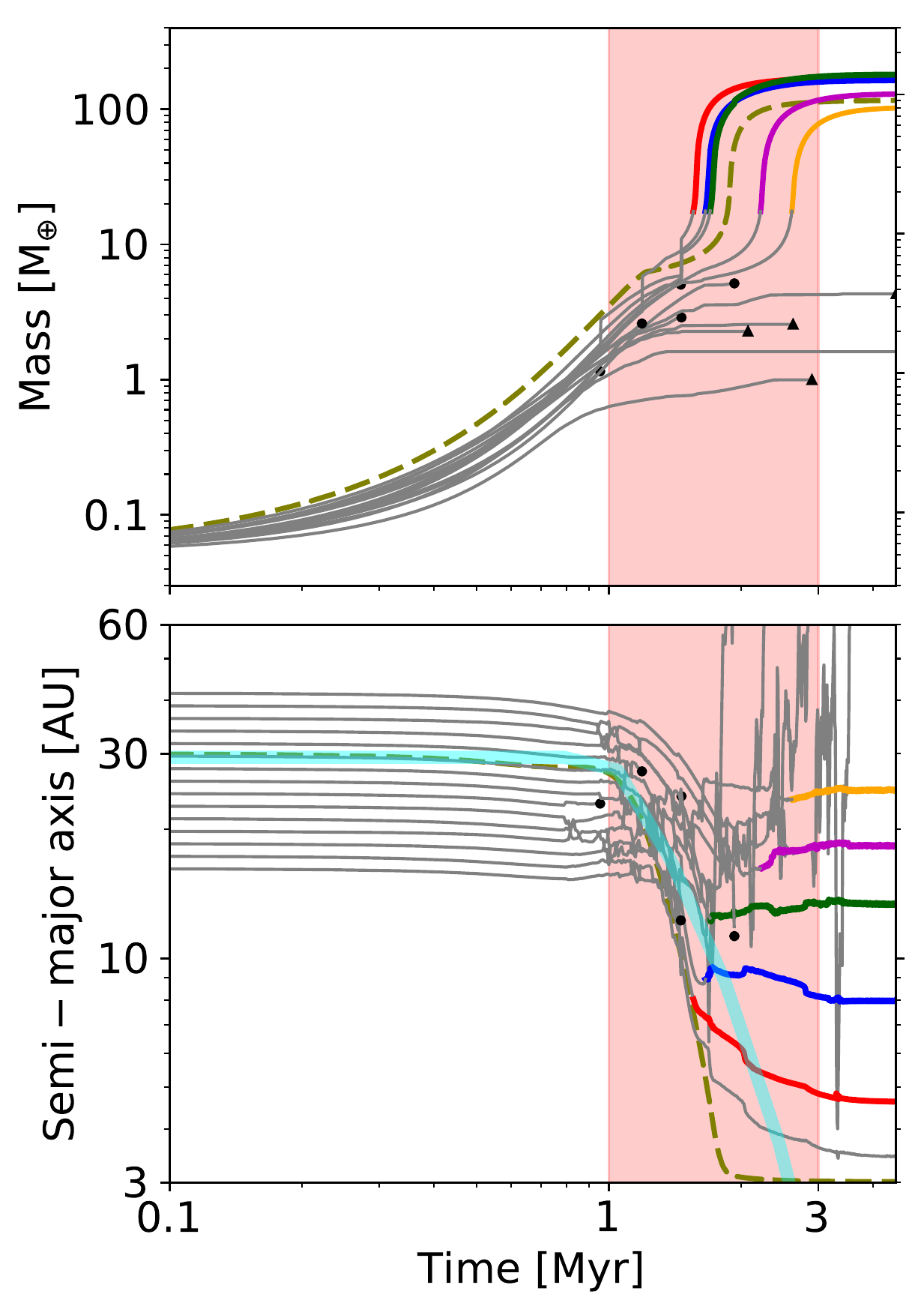}
  \includegraphics[scale=0.6, angle=0]{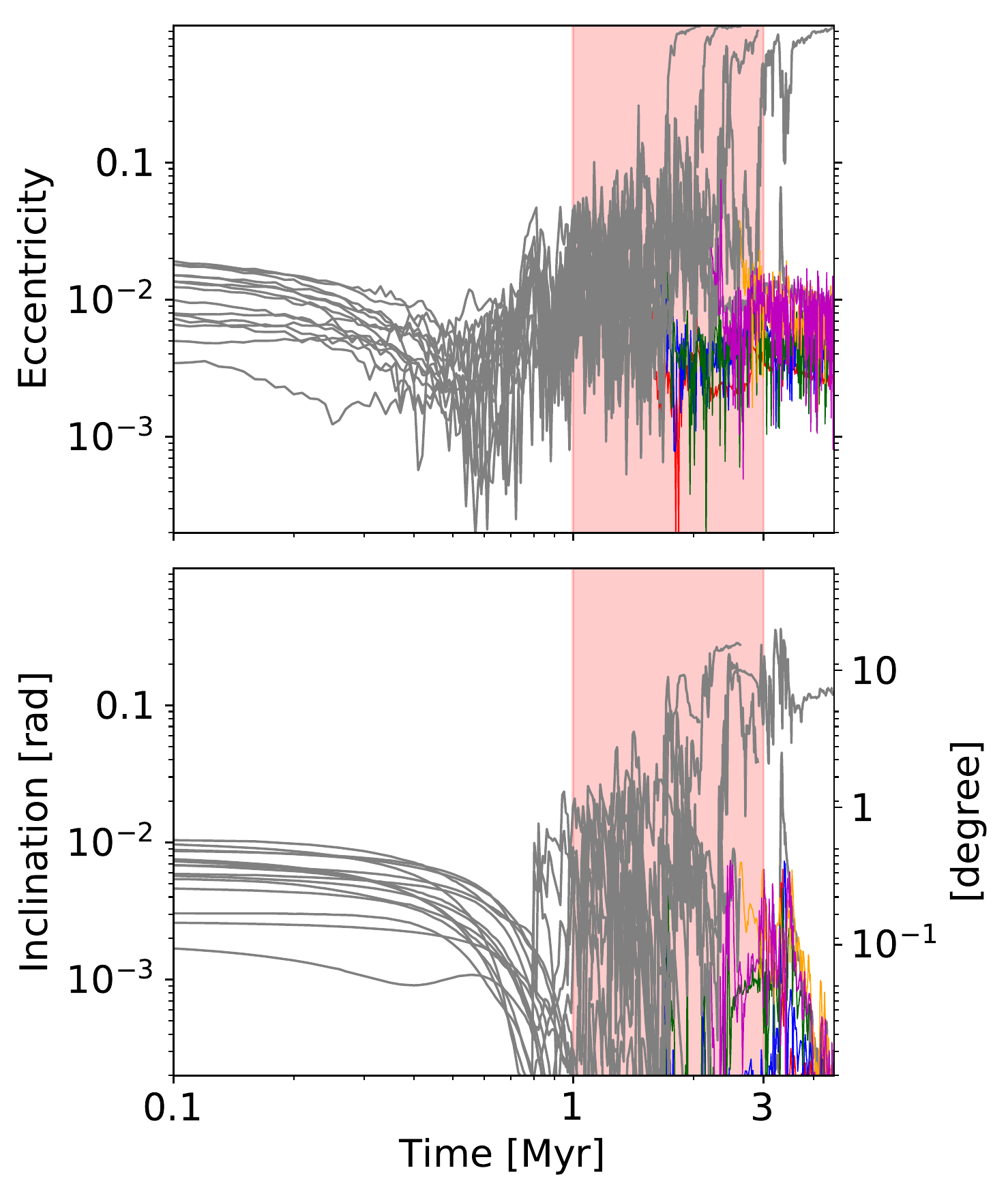}
       \caption{  
  Time evolution of planet mass (upper left),  semimajor axis (lower left), eccentricity (upper right) and inclination (lower right)  for fifteen $0.05\Me$ protoplanets, where they are initial distributed from $16$ AU to $43$ AU with a mutual separation of $15 \Rhill$.  The grey lines represent planets with $\Mp {<}\Mgap$, while the color lines represent the forming giant planets with $\Mp {>} \Mgap$. The circles and triangles indicate collisions and ejections, respectively. The thick cyan line is the transition radius and the yellow dashed line represents the growth and migration of  a single $0.05\Me$ protoplanet for comparison.  Multiple, wide orbit giant planets can efficiently form after $1.5{-}3$ Myr.  
    }
\label{fig:example}
\end{figure*}

\section{Parameter study}
\label{sec:parameter}
In this section we present a parameter study to investigate the influence of different parameters on forming giant planets. The parameters that we use in \se{result} is referred to as the fiducial case (\se{fiducial}). We vary one parameter in each subsection, including the birth location of protoplanets (\se{location}), the Stokes number of pebbles (\se{pebble}), the disk turbulent level (\se{turb}), the pebble flux (\se{flux})  and the gas disk depletion timescale (\se{depletion}).  For each case we perform eight sets of numerical simulations, with random orbital phase angles and $10\%$ variation of the semimajor axes of the planets.  The initial separations of these protoplanets are still adopted to be $15\Rhill$.  A summary of the numerical setups and statistical results are given in \tbs{setup}{parameter}.  
\Fg{parameter} shows a scatter plot of the final masses and semimajor axes of forming planets. The gap opening mass can be varied for cases and change with time.  We obtain $\Mgap(r{=}\rtran, t{=}0){=}18\Me$ for the fiducial case. Since we focus on the growth and migration of giant planets, only planets with masses higher than this gap opening mass are discussed in this section for the convenience of comparison.

  \begin{table}
    \centering
    \caption{Initial setup of the parameter study in \se{parameter}. The second, third fourth and fifth columns indicate the initial range of semimajor axes of planets,  the Stokes number of pebbles,  the local disk turbulent alpha, the pebble flux and the gas disk depletion timescale. In total fifteen $0.05 \Me$ protoplanets are simulated in the disk with an initial disk accretion rate of $10^{-7} \Msyr$. }
    \begin{tabular}{lclclclclclc|c|}
        \hline
        \hline
        Name       &   $a_{\rm p0}$   & $\taus$ &  $\alphat$  & $\dot M_{\rm peb} $  &   $\taud $\\ 
        & $\rm [AU]$&  & & $[\rm M_{\oplus}\ yr^{-1}]$ & $\rm [Myr]$ \\
      \hline
      \textit{Fiducial} & $[15,45]$ &  $0.01$ &  $10^{-4}$   & $1.5 \times10^{-4}$  & $0.5$  \\
       \textit{Ice line} & $[4,10]$  & $0.01$  & $10^{-4}$  &   $1.5 \times10^{-4}$ &  $0.5 $ \\
       \textit{High Stokes number}  &$[15,45]$& $0.1$  & $10^{-4}$  & $1.5 \times10^{-4}$  &  $0.5$  \\
     \textit{High disk turbulence} &$[15,45]$   & $0.01$ & $3{ \times} 10^{-4}$ & $1.5 \times10^{-4}$  &  $ 0.5$  \\
      \textit{Low pebble flux} &$[15,45]$   & $0.01$ & $10^{-4}$  &  $10^{-4}$ & $0.5$  \\
       \textit{Fast disk depletion} & $[15,45]$   & $0.01$ & $10^{-4}$  &$1.5 \times10^{-4}$ &  $0.25$  \\
          \hline
        \hline
    \end{tabular}
    \label{tab:setup}
\end{table}

  \begin{table*}
    \centering
    \caption{Statistical results for the parameter study in \se{parameter}. The second column gives the time when the first giant planet grow massive than $\Mgap$, while the third, fourth and fifth columns refer to the multiplicity,  mass and semimajor axis of the forming giant planets in the end of the simulation. Average value is provided in front and minimum and maximum values are given in brackets. Only giant planets with $M_{\rm p}{>}18\Me$ are considered here. }
    \begin{tabular}{lclclclclclc|}
        \hline
        \hline
        Name     & $t_{\rm p0}$   & $N_{\rm p}$   &  $ M_{\rm pf} $ &  $a_{\rm pf}$ \\ 
        & $\rm [Myr] $ &  & $[M_{\oplus}]$ &  $\rm [AU]$ & \\ 
      \hline
      \textit{Fiduical}  & $1.7$ $[1.4{-}1.9]$& $6$ $[4{-}8]$ & $ 99 $ ($20{-}208$) &$10 $ $[2{-}25]$ \\
      \textit{Ice line}  & $2.6$ $[2.2{-}3.1]$ & $3$ $[1{-}6]$ & $ 35 $ $[18{-}73]$ &$2$ $[1{-}4]$ \\
       \textit{High Stokes number}   & $2.1$ $[1.6{-}2.5]$ & $6$ $[3{-}8]$ &  $ 57 $ $[19{-}155]$ & $7$ $[2{-}24]$ \\
       \textit{High disk turbulence}   & $2.8$ $[2.8{-}2.9]$ &  $0.25$ $[0{-}1]$ & $ 89 $ $[84{-}94]$ & $8$ $[6{-}9]$ \\
         \textit{Low pebble flux}   & $3.5$ $[2.8{-}4.9]$ &  $0.63$ $[0{-}2]$ & $ 36 $ $[18{-}55]$ & $10$ $[5{-}15]$ \\
          \textit{Fast disk depletion } &  $1.8$ $[1.4{-}2.3]$ &  $2$ $[1{-}3]$   &  $ 56 $ $[19{-}104]$ & $12$ $[8{-}20]$ \\

          \hline
        \hline
    \end{tabular}
    \label{tab:parameter}
\end{table*}

\subsection{Fiducial case}
\label{sec:fiducial}

 \Tb{parameter} illustrates the mean (maximum and minimum) number, mass and semimajor axis of the forming giant planets that are massive than $\Mgap$. We find that the number of planets formed in each simulation ranges from $4$ to $8$, with a mean value of $6$.  It is worth noting that not all survived planets finally grow into gas giants. Compared to the growth of the single protoplanet, multiple protoplanets compete with each other for sharing the total pebbles in a disk. Mutual interactions among planets excite their eccentricities, which could casue either an increase or a decrease of pebble accretion, depending on the amplitudes of  their eccentricities \citep{Liu2018}. In addition, when the planets reach pebble isolation mass, they stop the inward drifting pebble flux \citep{Lambrechts2014b,Bitsch2018}. Therefore, any planets inside of those massive bodies cannot further accrete pebbles to grow their masses.

Among these forming planets, the highest mass is $208\Me$ and the average value reaches $99\Me$.  Regarding to the semimajor axis, the maximum and average values are $25$ AU and $10$ AU. It normally takes $1.7$ Myr, but this time can be as short as $1.4$ Myr.  Statistical results confirm that our scenario promotes the formation of multiple giant planets with larger orbital distances than the typical water-ice line location.

\subsection{Birth locations of protoplanets}
\label{sec:location}
In the fiducial case, planets are distributed in both sides of  $\rtran$  (${\sim}15{-}45$ AU). In this case we assume that planets are born near the water-ice line where the disk temperature is $170 \rm \ K$. The corresponding  location $r_{\rm ice}$ is at $7$ AU in the early gas-rich phase.  The planets are initially distributed at $4{-}10$ AU, with a mutual separation of $15 \Rhill$.

We find in \fg{parameter} that fewer giant planets form when the protoplanets are born near the water ice line (red triangle) compared to those formed at further out disk locations around $\rtran$ (black circle). The planets formed in this case also have lower masses and shorter orbital periods. The maximum planet mass is $73 \Me$ and the largest orbital distance is $4$ AU. 

 We note that  both the pebble accretion and the migration timescales are dependent on radial distance. 
 The type I migration timescale is given by 
   \begin{equation}
     \begin{split}
  \tau_{\rm mig}    =   \frac{1}{f_{\rm I}}  \left(  \frac{M_{\star}}{M_{\rm p}}  \right) \left(  \frac{M_{\star}}{\Sigma_{\rm g} r^2} \right) \left(\frac{h_{\rm g}^{2} }{\Omega_{\rm K}} \right)  
  \propto 
  \begin{cases}
 {\displaystyle  r^{-1/4}}    \hspace*{0.3cm}   [\mbox{vis}]  \\
    {\displaystyle  r^{8/7} }  \hspace*{0.5cm}   [\mbox{irr}],
\end{cases}
\label{eq:migration}
\end{split}
\end{equation}
and the growth timescale in the $2$D pebble accretion regime is given by 
   \begin{equation}
     \begin{split}
  \tau_{\rm PA,2D}   =   \left(  \frac{M_{\rm p}^{1/3} M_{\star}^{2/3}}{ 0.24 \dotMp}  \right) \eta \taus^{1/3}  
  \propto 
  \begin{cases}
 {\displaystyle  r^{-1/8}}    \hspace*{1.3cm}   [\mbox{vis}]  \\
    {\displaystyle  r^{4/7} }  \hspace*{1.5cm}   [\mbox{irr}].
\end{cases}
\label{eq:PAaccretion}
\end{split}
\end{equation}
The above radial distance dependence is derived from \eq{sigma}, \eq{aspect} and $\eta \propto h_{\rm g}^2$.

We find that  both the growth and migration are most efficient at $r_{\rm tran}$ in our model. 
When protoplanets form around the water-ice line, their growth is slightly slower compared to those at $\rtran$. Meanwhile, they also take longer time to migrate to $\rtran$. The disk already starts to dissipate before they reach $\rtran$, and therefore, the total collisions among these planets are fewer. As a result, we see that in this case the protoplanets grow to less massive giant plants with shorter orbital distances at a later time.  Similarly, we expect that the  growth  of planets would be suppressed when they form much further out compared to $\rtran$.

\subsection{Stokes number of pebbles}
\label{sec:pebble}

In this case we set the Stokes number of pebbles to be $0.1$, one order of magnitude higher than that in the fiducial case. The results are shown in \tb{parameter} and \fg{parameter} (orange square). We find that when the Stokes number of pebbles is higher,  planets generally have lower masses and slightly shorter orbital periods.

Since the radial drift velocity of pebbles increases with the  Stokes number, the pebble accretion accretion efficiency in the $2$D regime decreases due to the fast drifting pebbles. Thus, in this case the protoplanets grow more slowly by pebble accretion.  The migration of these lower mass planets is also weaker, further reducing the chance of planet-planet collisions. As a result, the overall mass growth is suppressed, from both pebble accretion and giant impacts. These less massive planets accrete gas more slowly and transition to the type II migration at later time. Therefore,  the masses and semimajor axes of the forming planets are lower and the growth time becomes longer compared to those in the fiducial case.  

\subsection{Disk turbulence}
\label{sec:turb}

We increase the disk turbulent viscosity by a factor of three and keep the rest parameters the same as the fiducial case. Compared to the fiducial case, we find in \fg{parameter} that the planet mass growth is strongly suppressed when the disk turbulent viscosity becomes higher (purple diamond). Finally, only two gas giant planets form out of eight simulated planetary systems (\tb{parameter}).

The strength of disk turbulence correlates with the scale height of pebbles. 
Pebbles are more vertically extended when the disk is more turbulent. This means that a smaller fraction of pebbles can be affected by the gravitational force of the planet and get accreted. Therefore,  the mass growth by pebble accretion becomes much less efficient in this case.  On the other hand, the pebble isolation mass is also higher in a more turbulent disk. The planets need to grow more massive to trigger rapid gas accretion. As a result, the massive gas giant planets can only form when the protoplanets undergo multiple collisions. This is why only a few  giant planets can form in this case.  In addition, based on \cite{Kanagawa2018}'s migration prescription, the gap opening mass that transitions from type I to type II migration is also higher in a more turbulent disk. The planet spends more time in the fast, type I migration rather than the slow type II migration. Thus, even in the optimistic case giant planets can form, they are likely to migrate further in compared to those in fiducial case.   
We also note that if the global $\alphag$ could affect local processes as mentioned in \se{method}, it then suggests that giant planets would be even harder to form.

\subsection{Pebble flux}
\label{sec:flux}

We also test a case when the disk pebble flux is reduced to $10^{-4} \Meyr$. Compared to the fiducial case, we also find in \fg{parameter} that the forming planets are less massive when the total pebble mass in the disk is lower (cyan hexagon). It takes longer time to grow giant planets as well (\tb{parameter}).

The pebble flux is crucial to form gas giant planets. Pebble accretion rate decreases with a decreasing  of disk pebble flux. Since fewer pebbles are available for accreting,  the growth in this case is slower compared to the fiducial case. One should note that the final mass of forming planet does not scale linearly with the pebble flux. While the pebble flux is only reduced by $30\%$ compared to the fiducial case, the average giant planet mass decreases by a factor of $2{-}3$.  The giant planet formation indeed requires a massive pebble disk.  
It is worth noting that the influence of pebble flux on the final planet mass is general, which have been found  in \cite{Bitsch2019} and \cite{Lambrechts2019} for both inner super-Earths/low-mass terrestrial planets and outer giant planets.  
 
\begin{figure}
 \includegraphics[scale=0.48, angle=0]{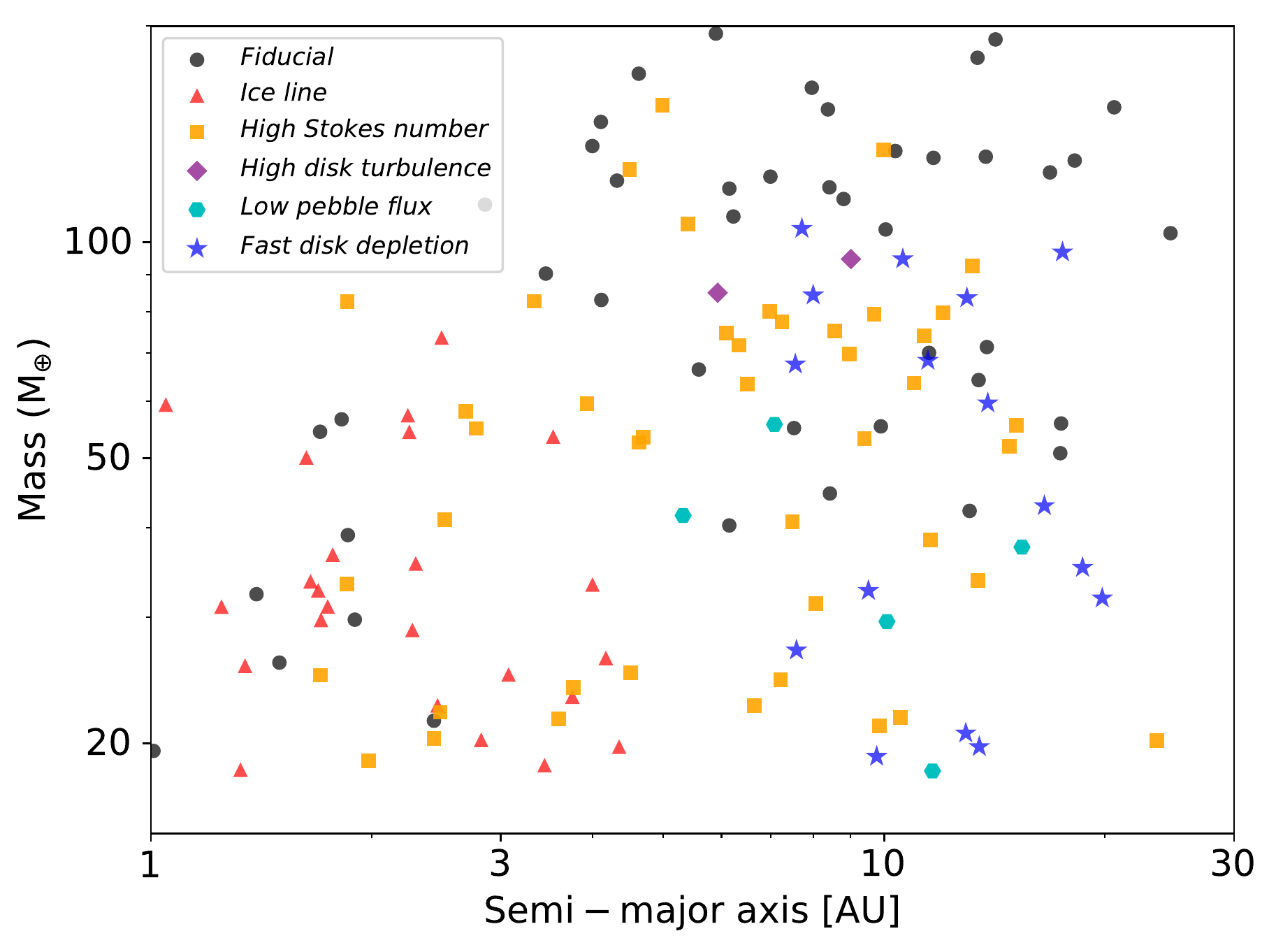}
       \caption{  Scatter plot of masses and semimajor axes of forming giant planets for the parameter study in \se{parameter}. The fiducial, ice line, high stokes number, high disk turbulence, low pebble flux and fast gas depletion cases are illustrated in  black circle, red triangle, orange square, purple diamond cyan hexagon and blue star, respectively. Giant planets with higher multiplicity, larger orbital distances and shorter formation time are more likely to form when they are born closer to the transition radius, the disk is more turbulent, the Stokes number of pebbles is lower, the pebble flux is higher and/or the disk gas removal is slower. 
    }
\label{fig:parameter}
\end{figure}
 
\subsection{Disk depletion timescale}
\label{sec:depletion}

We reduce the gas disk depletion timescale by a factor of two. In this situation on average only two giant planets form per system (\tb{setup}), much less compared to the fiducial case.  The forming planets have lower masses, but end up at larger orbital distances (blue star in \fg{parameter}).

On one hand, the faster dispersal disk contains less materials to be accreted by the planets. It results in an insufficient mass growth by pebble accretion.  On the other hand, less gas mass is left in the disk to transfer angular momentum to the planet, leading to a slower migration.  Therefore,  the protoplanets finally grow into in less massive planets with larger orbital distances in disks that dissipate gas more rapidly.

To summarise, giant planets with higher multiplicity, larger orbital distances and shorter formation time are more likely to form when they are born closer to the transition radius, the disk is more turbulent, the Stokes number of pebbles is lower, the disk pebble flux is higher and/or the disk gas removal is slower.

 \section{Discussion}
 \label{sec:discussion}
 
 \subsection{Number of protoplanets}
\label{sec:number}

\begin{figure}
 \includegraphics[scale=0.65, angle=0]{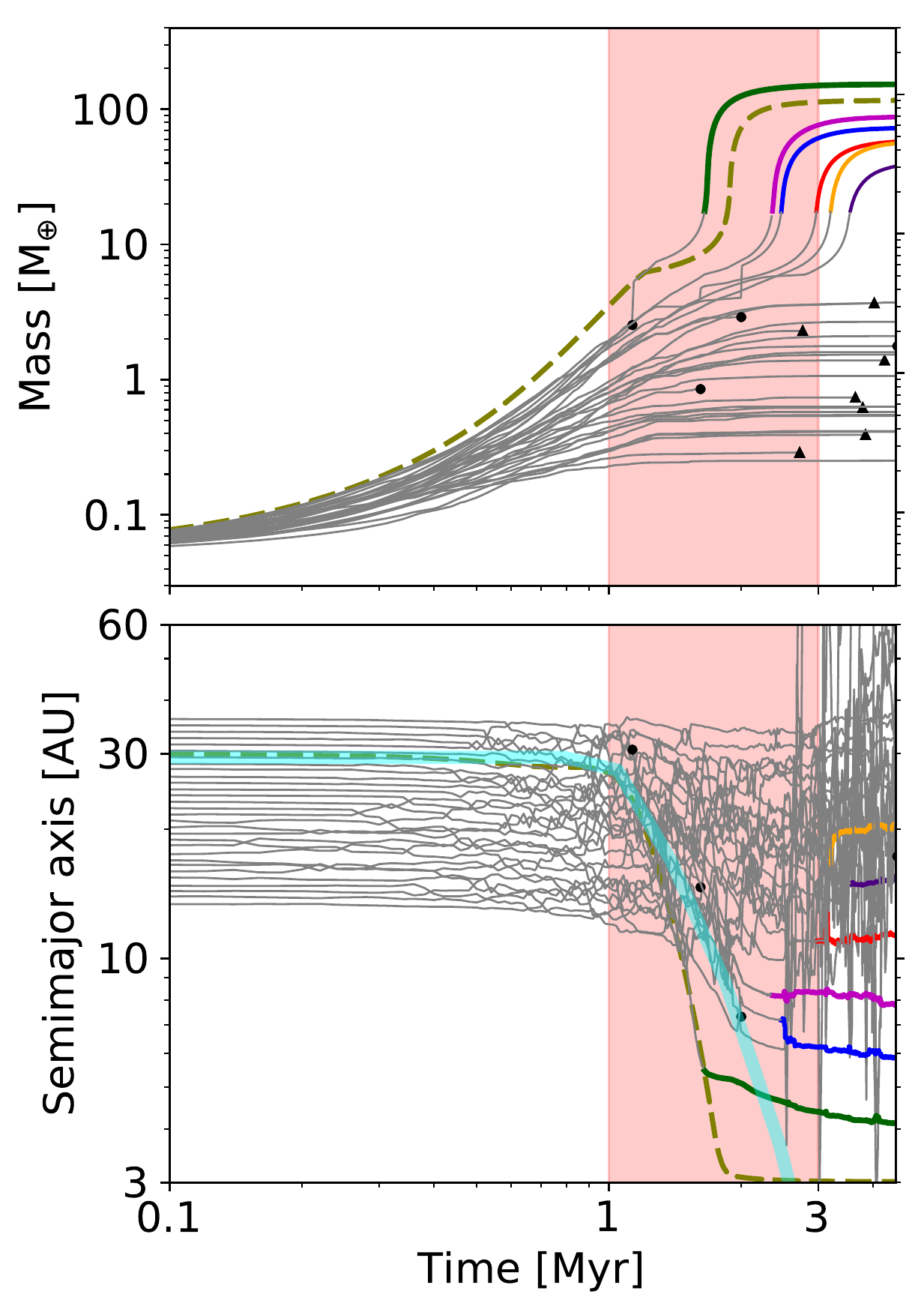}
       \caption{  
  Time evolution of planet mass (top) and  semimajor axis (bottom) for thirty $0.05\Me$ protoplanets, where they are initial distributed from $16$ AU to $43$ AU with a mutual separation of $7.5 \Rhill$.  The grey lines represent planets with $\Mp {<}\Mgap$, while the color lines represent the forming giant planets with $\Mp {>} \Mgap$. The circles and triangles indicate collisions and ejections, respectively. The thick cyan line is the transition radius and the yellow dashed line represents the growth and migration of  a single $0.05\Me$ protoplanet for comparison.  Compared to \fg{example}, an increasing number of protoplanets do not change the overall feature of forming giant planets.  
    }
\label{fig:N30}
\end{figure}

We choose a total number of $N{=}15$ protoplanets for the simulations presented in \ses{multiple}{parameter}. Rich dynamical features can be seen when we account for the multiplicity of the protoplanets. However, we still barely know that how many protoplanets are able to form in early disk. In order to explore the influence of number of protoplanets on the final outcome, we here test a case with $N{=}30$ and perform eight different sets of simulations with randomised initial conditions.  When the number of protoplanets increases by a factor of two, the number of final giant planets do not increase accordingly. We find that averagely $5$ giant planets form per system in this case, and the overall orbital features are quite similar to those shown in the fiducial case.

  \fg{N30} is an example illustrating the masses and semimajor axes evolution of the planets with $N{=}30$. 
 We emphasize here again that not all protoplanets  grow into giant planets. Interactions among planets stir up their eccentricities and inclinations, which affects the pebble accretion efficiencies. After the formation of a few giant planets, the system becomes dynamically hot. These giant planets strongly perturb and scatter the rests of the low-mass planets into more eccentric and inclined orbits, suppressing their further growth. Altogether, this self-regulated process limits the total number of gas giants.

% In the fiducial case we set the masses of protoplanets to be $0.05\Me$ at the beginning. We demonstrate that when the initial masses of protoplanets are lower, in order to grow into massive planets, we need a disk with high pebble fluxes.  It is obvious that the pebble accretion efficiency correlates with both the initial mass of the protoplanet and the total mass of pebbles in the disk.   Final planets grow less massive when the initial masses of protoplanets are lower and/or the disk pebble fluxes are lower.     A lower initial    flux  

\subsection{Collision vs scattering}
\label{sec:escape}
The highest relative velocity that two planets can reach is the surface escape velocity $v_{\rm esc} {=} \sqrt{2 GM_{\rm p}/R_{\rm p}}$ where $G$ is the gravitational constant and $R_{\rm p}$ is the physical radius of the primary planet.  Whether a close encounter between two planets leads to a scattering or a collision is determined by the ratio of the surface escape velocity $v_{\rm esc}$ and the escape velocity of the planetary system (${=}\sqrt{2}v_{\rm K}$ where $v_{\rm K}$ is  the Keplerian velocity at the planet location) \citep{Goldreich2004}.  We quantify this by     
\begin{equation}
      \begin{split}
\Lambda^2 & =  \frac{v_{\rm esc}^2}{2 v_{\rm K}^2} = \left(\frac{M_{\rm p }}{M_{\star}} \right)\left(\frac{a_{\rm p }}{R_{\rm p}} \right) \\
& \simeq 2 \left(\frac{a_{\rm p }}{30 \AU} \right)  \left(\frac{M_{\rm p}}{M_{\oplus}} \right)^{0.5} \left(\frac{M_{\star }}{M_{\odot}} \right)^{-1},
 \end{split}
 \label{eq:esc}
\end{equation} 
where we use the mentioned mass-radius relation from \cite{Lissauer2011} to derive the latter equation.    
When $\Lambda{\gg}1$ (the escape velocity is much larger than the Keplerian velocity), the outcome during planet-planet encounters are likely to be ejections rather than collisions. On the other hand, when $\Lambda{<}1$, merger between the two planets will be the favoured outcome.  
It is worth pointing out that the above analytical estimation is based on a gas-free condition. However, when the gas disk is present, the velocity dispersion of the planets is also damped by disk gas.  The actual relative velocity between approaching two planets is generally smaller than $v_{\rm esc}$. Nonetheless,  the above scalings indicate that planets tend to be ejected when planets are more massive and/or have larger orbital distances.

For the low-mass planets of $M_{\rm p}{<} \Miso$, when encounters for such planets occur at  $r{\lesssim} \rtran$, collisions are still preferred than ejections ($\Lambda {\lesssim}1$). However, ejections is more favoured for massive gas giant planets ($\Lambda {\gg}1$).  Such a feature can also be seen in \fg{example}. At early times low-mass protoplanets  collide to grow their masses (circles). Once the planets accrete substantial gaseous envelopes and become giant planets of ${\gtrsim}100 \ M_{\oplus}$, they tend to eject the approaching low-mass planets instead (triangles).    

Similarly, we also expect ejections would be dominated when planets are located at very large orbital distances (\eg , $r_{\rm p}{\gtrsim}100$ AU),  and thereby the core growth is only led by pebble accretion.  \cite{Ormel2018} found that pebble accretion efficiency also decreases with radial distance. Therefore, this formation channel is unlikely to grow massive giant planets at very wide orbits.

\subsection{Comparison with other studies}
In our disk model the location of the planet trap is at the transition radius, which separates two disk heating mechanisms.  In addition, several other studies proposed that planets can be trapped at the different disk locations due to dust sublimations and opacity transitions \citep{Bitsch2013,Bitsch2015,Baillie2015}.  Although the detailed disk models can be different, we would like to point out that many studies consistently suggested such planet traps are sweet spots for the growth of planet cores and promote the formation of gas giant planets \citep{Lyra2010,Hellary2012,Cossou2014,Liu2015}.

Recently, several studies incorporated pebble accretion into N-body code to investigate the giant planet formation \citep{Levison2015,Matsumura2017,Bitsch2019}.  \cite{Levison2015} successfully reproduce the architecture of  giant planets in Solar System without taking into account of planet migration.  When including migration,  \cite{Matsumura2017} however found that distant giant planets are difficult to form. This is partly due to the fact that they choose a classical type II migration prescription (their Eq. 26). It means that the migration rate is independent of the planet mass in the disk-dominated regime. However, \cite{Kanagawa2018} found that the  migration rate decreases with the planet mass when the planet massive enough to open a deep gap.  We adopt \cite{Kanagawa2018}'s prescription, which leads to less significant planet migration in our cases compared to \cite{Matsumura2017}.  \cite{Bitsch2019} explored the influence of pebble flux and found that in order to compete the inward migration, the (multiple) giant planet formation requires a sufficiently high pebble flux, which is consistent with our findings in \se{flux}.

\section{Conclusions}
\label{sec:conclusion}

In this paper we explore a scenario for the early formation of multiple, distant giant planets. Such planet populations are observed in radial velocity and microlensing surveys, as well as inferred from the substructures exhibited in young protoplanetary disks.  Based on the planet formation model of \cite{Liu2019b},  we performed N-body simulations to study the growth and migration of a large number of protoplanets in disks during gas-rich and depletion phases. The physical processes that we take into account are pebble accretion onto planet cores, gas accretion onto planet envelopes, planet-planet interactions/collisions, type I and type II planet migration.
The initial physical properties of the planets and disk are given by a set of parameters $a_{\rm p},\ \taus,\ \alphat,\ \dot{M}_\mathrm{peb},\ \mathrm{and}\ \tau_\mathrm{dep}$ which will govern the outcome of our simulations. We conduct a parameter study to investigate the influence of  these parameters. In turn, we show that there exists an optimal case where our model can  produce multiple giant planets at distances between three and a few tens of AU in a short time span of a few Myr.

The key findings are summarised as follows.
\begin{itemize}
\item[--] For the single protoplanet growth case, the core grows only by accreting pebbles. The final location of the planet is determined by a competition between pebble accretion and inward migration. Finally, the protoplanet  can grow into a gas giant planet at an orbital distance ${\lesssim} 3$ AU (\fg{mr}).
\item[--]When multiple protoplanets co-exist,  they undergo convergent migration to the transition radius ($\rtran{=}30$ AU at early phase). A temporary trapping of planets at $\rtran$ causes the orbits of the planets frequently overlap with each other and eventually planet-planet collisions.  In this case the rapid growth is because of a combined of pebble accretion and direct planet mergers. This also results in a quick transition from the fast, type I migration to the slow, type II migration. Therefore, the giant planets form early and end up at larger orbital distances compared to the case when only single protoplanet is considered. Furthermore, the formation of early giant planets promotes the subsequent growth of massive planets with the orbits exterior to them (\fg{example}). 
\item[--]  Massive, distant giant planets are less likely to form when the initial protoplanets are born closer-in, the Stokes number of pebbles is higher,  the disk is more turbulent, the pebble flux is higher  and/or disk gas depletion is faster (\tb{parameter} and \fg{parameter}).  
\end{itemize}

%\revise{Our model does not attemp to sove all exisiting planet formation puzzles. For instance, we are still fail to explain very distant giant planets which frequetly osbervsed and interpretaed  from ALMA surveys. Alternatively, our proposed scneario provides a tentitave explaination, which has potential to link to some aspections of planet. }

 We note that we opted for a simplistic distribution of planetesimals in our simulations, only tracking the growth of the most massive protoplanets produced by streaming instability in the disk within a limited semimajor axis range. Such a distribution serves the  purpose of producing multiple, distant giant planets, which is the goal of this paper. In future work, we aim to implement a more realistic distribution of protoplanets to study whether or not our model can reproduce additional architecture in planetary systems.
This study can also be extended by generating distributions of model parameters in a Monte Carlo manner. The resulting planet populations can then   be statistically compared with the inferred planets from microlensing surveys \citep{Suzuki2018} and  ALMA disk observations \citep{Nayakshin2019,Ndugu2019}.

\section*{Acknowledgements}
We thank Doug Lin, Feng Long for fruitful discussions, and Anders Johansen, Michiel Lambrechts for proofreading the manuscript and providing helpful comments. We also thank the anonymous referee for their useful suggestions. 
B.L. is supported by the European Research Council (ERC Consolidator Grant 724687-PLANETESYS) and the Swedish Walter Gyllenberg Foundation.  J.W. thanks Anders Johansen's financial support for his summer project.
M.O. is supported by JSPS KAKENHI Grant Numbers 18K13608 and 19H05087.

 \appendix
 \section{Type I migration prescription}
\label{ap:torque}

 Type \Romannum{1}  migration prescription is adopted from \cite{Paardekooper2011}. The type I torque includes the differential Lindblad torque $\Gamma_{\rm L}$, the barotropic part of the horseshoe drag $\Gamma_{\rm hs, baro} $ and linear corotation torque $\Gamma_{\rm c, lin, baro}$,  entropy-related the horseshoe drag $\Gamma_{\rm hs, ent}$ and linear corotation torque  $\Gamma_{\rm c, lin, ent}$. Each component of the torque can be found in   Eqs. $3{-}7$ of \cite{Paardekooper2011}. Here the non-linear horseshoe drag and linear corotation torque are together called corotation torque $\Gamma_{\rm c}$.

The total type I torque is given by
\begin{equation}
\begin{split}
\Gamma_{\rm I}  &  =  f_{\rm I}\Gamma_0  =  \Gamma_{\rm L} +  \Gamma_{\rm c} =  \Gamma_{\rm L} +  \Gamma_{\rm hs, baro} F(p_{\nu})G(p_{\nu})\\
& + (1 - K(p_{\nu}))  \Gamma_{\rm c, lin, baro}  +   \Gamma_{\rm hs, ent} F(p_{\nu}) F(p_{\chi}) \sqrt{G(p_{\nu})G(p_{\chi})} \\
& + \sqrt{(1-K(p_{\nu}))(1-K(p_{\chi}))  } \Gamma_{\rm c,lin, ent}, 
\end{split}
\label{eq:torque1}
\end{equation}
where $\Gamma_0$ is the normalized type \Romannum{1}  torque, the adiabatic exponent $\gamma$ is adopted to be $1.4$, $F(p)$, $G(p)$, and $K(p)$  are fitting smooth functions that describe the saturation of corotation torque (their Eqs. 23, 30, 31),  and $p_{\nu}$, $p_{\chi}$ are the saturation parameters related with viscous and thermal diffusion: %(see their Eqs. 19 and 40) 
\begin{equation}
p_{\nu} = \frac{2}{3} \sqrt{ \frac{ \Omega_{\rm K} r^2 x_{\rm s}^3 }{2 \pi  \nu } }, p_{\chi} = \frac{3}{2} p_{\nu}  P_{\rm r}^{1/2}, 
\label{eq:pp}
\end{equation}
where  $P_{\rm r} {=} \nu / \chi  $ is the Prandtl number. 

In the inner optically-thick disk viscously heated regions,  we assume that the turbulent and thermal diffusion are comparable ($p_{\chi}{=}p_{\nu}$) for simplicity, and thus Prandtl number is of order unity. Planet with certain mass can migrate outward, depending on $\alpha_{\rm t}$ and $h$ (see \eq{mopt}).
% For the further out disk region in optical thin regime, $\chi$ can larger than $\nu$ and  the entropy-related corotation torque would  diminish \citep{Paardekooper2011}. However, the disk opacity at $30{-}60$ AU is around $3 \ \rm  cm^{2} \ g^{-1}$. The optically thin criterion requires $\Sigma_{\rm g} {<}0.3 \rm \ g \ cm^{-2} $, corresponding to $\dot{M}_{\rm g} {<}10^{-9} \Msyr$ (\eq{sigma}). This only occurs after $3$ Myr for our  fiducial model, when all giant planets have already formed. Thus, we do not expect that our results are largely  affected by the above simplification. Importantly, in the outer irradiation-dominate disk regions, planets always migrate inward, no matter the entropy-related torque is saturated or not. We check the magnitude also varies by $30\%$. Our approximation is still reasonably good. 
In the outer irradiation-dominated region, the entropy-related corortation torques vanish ($\Sigma_{\rm g,irr} {\propto}r^{-15/14}, T_{\rm irr}{\propto}r ^{-3/7}$). \Eq{torque1} reduces to the isothermal case, and planets only undergo inward migration.

 We do not account for the diminishing of the corotation torques due to an increasing eccentricity and inclination of the planet \citep{Bitsch2010}. This could weaken the outward migration in the multiple-protoplanet cases. On the other hand, we also neglect the stochastic torques due to the fluctuation nature of disk turbulence \citep{Ogihara2007,Baruteau2010}. Notably,  \cite{Pierens2013} found that when considering this stochastic torque, resonant chains are more easily disrupted and the planet-planet collisions are enhanced.  We will investigate the influence of these two effects  in a separate paper.  

Based on our disk model and the type \Romannum{1}  migration prescription, we show the  migration map of the type \Romannum{1}  coefficient $f_{\rm I}$ in \fg{map}.  The dashed line represents the optimal planet mass for outward migration in \eq{mopt}. As planets grow to ${\sim}1{-}6 \Me$, they can migrate outward and retain close to $r_{\rm tran}$.  Combining \fg{map} and \fg{example} We can have a better understanding of planet migration behavior.

\begin{figure}
 \includegraphics[scale=0.55, angle=0]{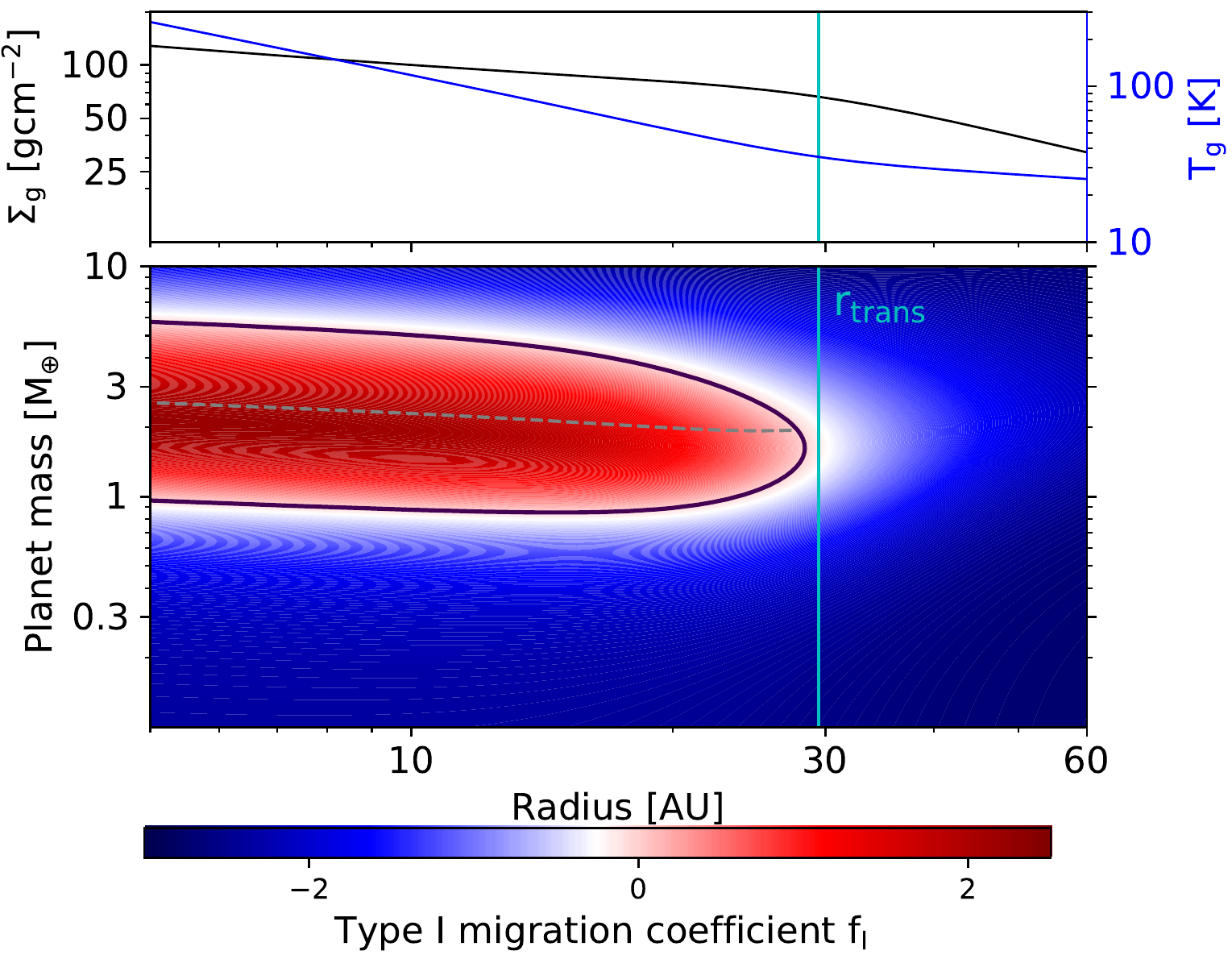}
       \caption{  Top:  gas surface density and disk temperature as functions of disk radius.
       Bottom: type I migration coefficient $f_{\rm I}$  as functions of the planet mass and disk radius.  The red (blue) indicate that migration is outward (inward), the black line refers to the zero-torque location, and the cyan line is the location of the transition radius $r_{\rm tran}$. The optimal mass in \eq{mopt} for planets of outward migration in the viscous heated region is shown in grey dashed line.  The adopted disk parameters are $\alpha_{\rm g}{=}10^{-2}$, $\alpha_{\rm t}{=}10^{-4}$, $\dot M_{\rm g} {= }10^{-7} \rm \ M_{\odot}yr^{-1}$. We note that in this illustration only type I migration coefficient is presented, and the transition to type II when the planet approaches the gap-opening mass is neglected. 
    }
\label{fig:map}
\end{figure}

\bibliographystyle{mnras.bst}
\bibliography{main.bib}

\label{lastpage}

\end{document}